\documentclass[english,aps,pra,reprint,showpacs]{revtex4-1}   
\usepackage[T1]{fontenc}	
\usepackage[latin9]{inputenc}	
\usepackage{geometry}		
\usepackage{graphicx}
\usepackage[above,below]{placeins}	
\usepackage{times}
\usepackage{csquotes}
\begin{document}

\title{Student self-assessment and reflection in a learner controlled environment}
\author{Jeffrey A. Phillips}
\affiliation{Department of Physics, Loyola Marymount University, 1 LMU Drive; MS-8227, Los Angeles, CA, 90045}


\begin{abstract}
Students who successfully engage in self-regulated learning, are able to plan their own studying, monitoring their progress and make any necessary adjustments based upon the data and feedback they gather.  In order to promote this type of independent learning, a recent introductory mechanics course was modified such that the homework and tests emphasized the planning, monitoring and adjusting of self-regulated learning.  Students were able to choose many of their own out-of-class learning activities.  Rather than collecting daily or weekly problem set solutions, assignments were mostly progress reports where students reported which activities they had attempted, self-assessment of their progress and plans for their next study session.  Tests included wrappers where students were asked to reflect on their mistakes and plans for improvement.  While many students only engaged superficially the independent aspects of the course, some did demonstrate evidence of self-regulation. Despite this lack of engagement, students performed as well as comparable student populations on course exam and better on the Force Concept Inventory.
\end{abstract}

\pacs{01.40.Fk, 01.40.gb}

\maketitle

\section{Introduction}
In most introductory physics courses, students are assigned required weekly problem sets that consist of five to ten problems (or exercises) from the textbook. While this structure provides students opportunities to practice applying key concepts to new situations, it does guarantee learning. By providing all students with the same list of practice problems, an instructor is not providing personalized practice. 

While it could be argued that the best scenario would be for the instructor to provide personalized practice for each student, this is often impractical and sends the message to students that they are dependent on an authority for their learning.  While instructors are in the position to offer feedback and suggestions to students, the students need to assume some responsibility for their own learning. Ideally, a student would employ metacognition and engage in self-regulated learning (SRL), which broadly describes a process by which a learner plans his/ her task, monitors the work and thinking during the task and makes adjustments based upon the data gathered \cite{zimm}. A quote, typically attributed to John Dewey, best expresses this need for SRL: "We don't learn from experience. We learn from reflecting on experience."

To examine whether or not personalized learning and scaffolded self-regulation could impact student performance, a physics course utilized ideas from learning-controlled instruction (LCI) \cite{merril}. Students were asked to select their own practice problems, with only a suggested list provided. To shift students' practice and attitudes about learning, they were asked to also engage in some guided self-regulation via prompts provided in homework reports. Each of these features is described below, as are some of the results, which show that very few students shifted their views or practices significantly. What the reports and associated data collection did do is paint a clearer picture of what students are doing in the out-of-class practice and provide a glimpse of the impact homework without reflection has on student learning.

\section{Methods}

\subsection{Context}

A section of an algebra-based mechanics course, which is typically taken by junior life science majors, was modified to included flexible out of class practice. The section began with 25 students, with one of them withdrawing prior to the end of the semester. The students' incoming GPA, 3.38; pre-instruction FCI score \cite{fci}, 9.3; and scientific reasoning (as measured by the Classroom Test of Scientific Reasoning (CTSR) \cite{lt}), 75.2\%, were statistically identical to prior sections'.

The course included many traditional components of an introductory physics course- a weekly two-hour lab and three 50-minute lecture sessions that included four in-class tests and final exam. Homework comprised 20\% of the students' final grade and consisted of two main types. In the first, students did short, almost daily, "warm-up" assignments from the textbook's accompanying workbook \cite{knight}. These activities were aimed at practicing fundamental skills in problem-solving and refining their understanding of physical models. Students were given credit if they came to class having attempted the activities; correctness was not a criteria. At the start of each class, the students would discuss their results and questions in groups. The class would then build on of the questions raised by students and any extensions or variations that the instructor added. 

The second type of homework was practice solving word problems, where students applied physics concepts and models to new, real-world scenarios. These problems came from the textbook as well as other resources, such as the University of Minnesota context rich archive \cite{knight2, minn}. Rather than assign a single set of practice problems each week, students were provided with a list of 50-75 suggested problems in each of the four units. These were organized by learning outcome (content focused) and sorted by difficulty. This made it easier for students to locate the practice problems that were appropriate for them at that time. Much like other previous learning controlled homework systems, this one allowed students to pick the problems that they wanted to practice \cite{bao}. Students were not required to attempt any set number or type of problems, they were free to choose their practice problems. Rather than submitting the solutions, students were required to submit reports in which they describe their planning, monitoring and adjusting.

\subsection{Homework Reports}
\begin{figure*}
\includegraphics[width=0.9\linewidth]{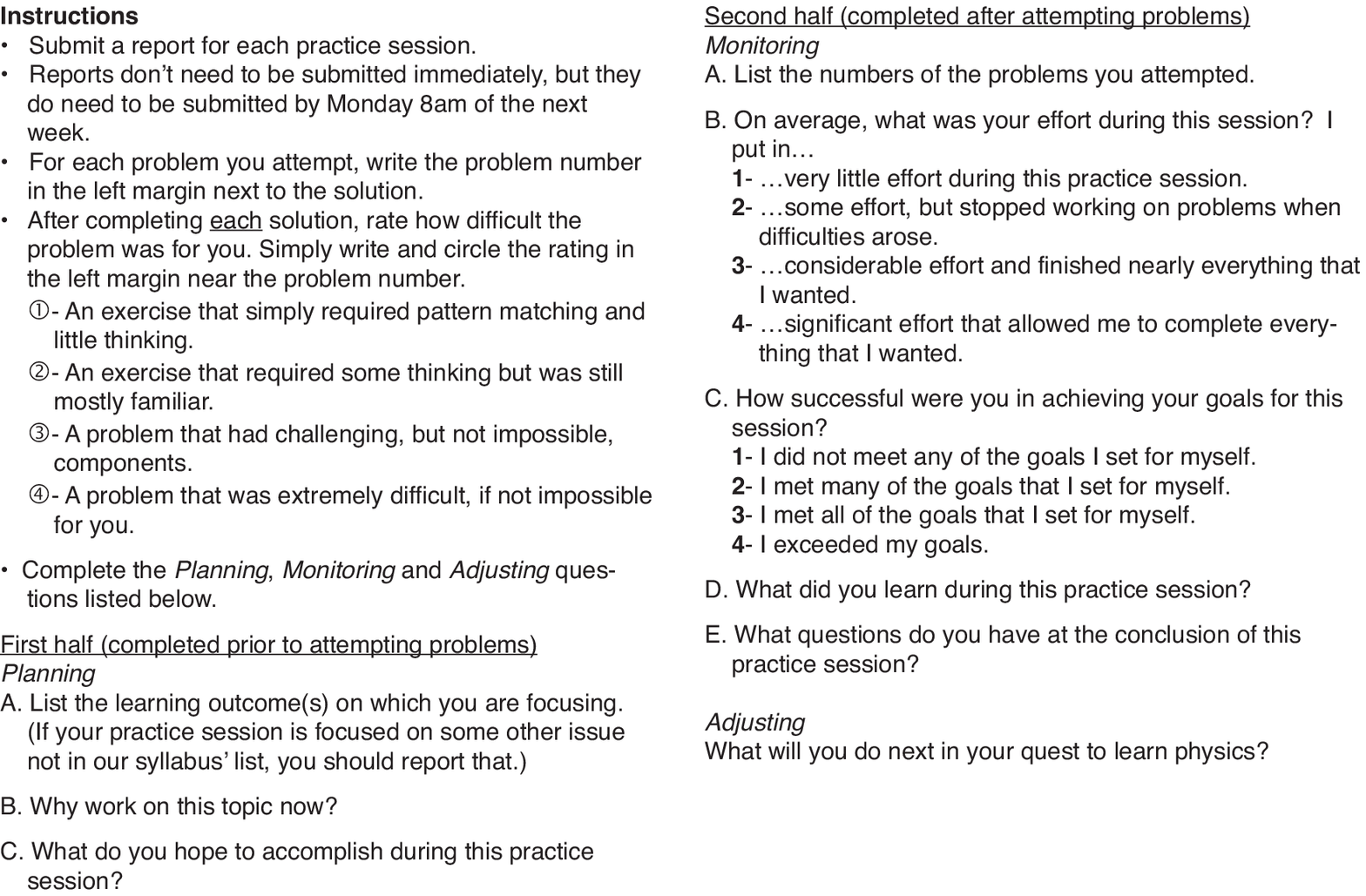}
\caption{Homework report instructions provided to students.\label{fig1}}
\end{figure*}

To collect information about what problems students were selecting and the degree to which they are engaging in self regulated learning, students were asked to submit a report describing their practice and its effectiveness each time they worked on physics problems outside of class. The reports had two components- self evaluation of problem difficulty and scaffolded reflection questions. The questions were divided into two halves- one that was to be completed when they were beginning their practice and the other at the end. The first half included prompts that asked students to describe their goals for that particular practice session (Planning). (Fig. \ref{fig1})  The second half asked students to reflect on the effectiveness of their practice (Monitoring) and articulate a plan for what their next practice will be (Adjusting).

The report was not too cumbersome as it could be completed on a single hand-written page. Complete sentences weren't required. Because of a desire to have the students solve their homework problems with a think-aloud protocol, each student was given a smartpen that electronically captured audio and penstrokes. (http://www.livescribe.com) The recordings made by the smartpens are essentially real-time movies of what was written and said. Students used this same technology to capture the homework reports. Students would write, or speak, out their responses, then once they synced the pen with their computer (via USB), the reports would automatically be emailed to the instructor.

At a minimum, one report needed to be submitted each week. If a student did not have the time to work on any practice within a given week, they could submit a report that described their plans for their future studying. As long as the Adjusting section was completed, students were given credit for submitting a report, no matter its quality. 

\section{Results}
\subsection{Class-wide}
Reports were required in ten of the course's fifteen weeks. In any week, roughly 5/8 of the students submitted complete reports, 1/8 of the students submitted only the Adjusting portion and 1/4 of the students failed to submit anything. All but two students consistently submitted only one report per week, usually in the 24-hour window before the deadline. Some students admitted to doing unreported practice immediately before a test, but generally it seemed that students only sat down to do practice problems once a week. In each session, they would report an average of three practice problems, with the maximum number being fifteen.

Most of the report were superficial, with the most common description of students' motivation for solving problems was to "do practice."  The purpose of doing the practice was the practice itself, not any specific learning objectives or shortcoming that they were looking to address. Despite the guided prompts, and class discussion about learning, most students saw (at least partially) the practice as the goal rather than the process by which a goal is achieved.
A few students did offer some more specific motivation:

\begin{itemize}
\item{\textit{After the exam I realized that even though I understood the theory behind problems, I need to work on challenging application word problems.}}
\item{\textit{some of the examples in class lost me; work on understanding signs better}}
\item{\textit{be able to solve problems in a timely manner without outside help}}
\end{itemize}

Just as superficial as the motivation (Planning) was, so was the reflection (Adjusting). Most students simply listed "do more practice" as their next steps in studying physics. Some did articulate more specific steps, which often involved seeking help from the instructor. The specificity or quality of the responses on the reports did not correlate with grades or problem-solving proficiency.

Neither was there a correlation between the number, or type, of practice problems reported and test grades. Some students did admit to doing some unreported practice that could account for the lack of correlation. Another explanation though could simply be that the practice is not having an impact on the students' learning. If they are not engaging in some metacognitive thought, they simply may not be learning from the practice that they are doing.
On an end-of-the-semester questionnaire, nearly all of the students spoke very highly about being able to choose their own practice problems. 

\begin{itemize}
\item{\textit{Yes, I thought it was helpful because you could focus on certain areas that you needed to work on more.}}
\item{\textit{Yes, because it allowed me to focus on problems that I specifically needed help on. I could concentrate my focus to one area at a time if that is what I needed. It's nice to have homework tailored to each individual students needs.}}
\item{\textit{Yes, I really did like this because it meant I did not have to waste my time on a  bunch of Level I or II's if I could easily do them, and I could focus on the more challenging problems.}}
\end{itemize} 

The students understood the value in tailoring their practice to their own needs. What they said that they did not enjoy or benefit from were the homework reports. 

\begin{itemize}
\item{\textit{I found the homework report somewhat unnecessary because I personally would have completed the suggested problems with or without the required reports.}}
\item{\textit{They [the reports] took too long.}}
\item{\textit{I never really understood the use of the before and after questions.}}
\end{itemize}

From these responses and their actual homework reports, it is very clear that most students did not engage in self-regulated learning. Even though they liked the idea of being in control of their learning and selecting the practice problems, they did not take complete control of that process.

\subsection{Case Study}
Isaac was perhaps the one student who showed and reported a noticeable amount of self-regulated learning, much of which he described as new for himself. From the beginning of the semester, he was somebody who was at risk for struggling in physics. Entering the course, he had a GPA that was 0.3 lower than the class average. On the CTSR, he had a score that indicated a lack of formal operational thinking (63\%). On the first three in-class tests, his scores were 18-24\% below the class average. Despite these difficulties, Isaac improved his performance by the fourth in-class test, scoring only 6\% below the class average and professing a change in his habits and views of learning.

Throughout the semester, but especially between the third and fourth test, Isaac described engaging in self-regulated learning \cite{phillips3}. Unlike his peers, he often described motivation for working on physics problems that was rooted in his past performance. 

\begin{itemize}
\item{\textit{To better understand free fall motion because I have been struggling a bit}}
\item{\textit{I hope to understand impulse/ momentum and how to better define my systems}}
\end{itemize}
 
In the open-ended questionnaire at the end of the semester, Isaac not only professed his appreciation for the learner-selected homework, he also commented on the usefulness of the weekly reports:

\blockquote{\textit{[The reports] made homework much more bearable and felt as if it had a purpose because you are the one that had set the goal for yourself. The monitoring portion of the homework report helped me alot because it made me think if I had any questions and if I didn't why was that, was I not doing hard enough problems or do I understand the concepts?}}

The insight that he has into interpreting the Monitoring portion of the report seems to indicate that he actually took the reports to heart and actively engaged in the process. The vast majority of students would usually say something to the effect of "I have no questions" in the Monitoring portion. Isaac was the only person who seemed to view the lack of questions as a problem. While not all students asked the instructor questions in their reports, enough did that it would seem that embarrassment wasn't the main reason for a lack of stated questions.

In addition to the improved test scores, Isaac also showed improvement in his physics conceptual understanding that was better than expected. His FCI normalized gain was much higher than students with similar scientific reasoning scores (0.70 vs 0.34) \cite{phillips1}. Perhaps because of his self-regulated learning, he exceeded expectations.

\section{Conclusions}
The ability to engage in self-regulated learning is key to students' success in a wide variety of courses as well as life after graduation. This autonomous learning can be viewed as one of the primary components to lifelong learning. Once learners take an active role in planning, monitoring and adjusting their learning, they no longer are reliant on instructors. The skills and values behind this independent learning are core to a liberal education \cite{aacu}.
   
Self-regulation skills, and/ or related values, are ones that also might transfer to smaller scale tasks such as solving word problems. In these shorter, more focused tasks, students also benefit highly from planning their solution before beginning the calculations, monitoring their thinking throughout so they may identify errors and adjusting any plans to correct errors \cite{schoenfeld, phillips2}.  

For at least one student, Isaac, the flexible homework structure and reports appeared to have promoted self-regulated learning behaviors, which impacted his class performance and attitude. For the rest of the class, it was unclear if there was a significant impact. At the very least, the flexible homework assignments did not degrade the class performance on in-class tests and final exam. While the tests vary from year to year, the final exam is unchanged making a direct comparison possible. There was no statistically significant difference between the exam scores of this section and those of prior two years, in which there were required weekly problem sets. On the FCI, this section actually showed a statistically significant larger normalized gain than in prior two years (0.57 vs 0.45, \textit{p} <0.05). 

Given the students' performance with only doing a small number of practice problems out of class each week, instructors may wish to reconsider the size of their assigned problem sets. A learner controlled homework system appears to provide comparable learning that one that is completely instructor directed. Also, moving away from collecting problem solutions, which didn't degrade student performance, could be a route for instructors who wish to avoid the plagiarism issues that plague many introductory courses \cite{chron}. 

Based on their reports, it seems that most students did not engage in self-regulated learning. Without this reflection, out of class practice does not produce significant learning. Given the superficial reflections included in the homework reports, it seems likely that most students do not carefully plan, monitor or adjust their practice. If this is the case, it is plausible that homework sets of any size, when not accompanied by reflection, won't have significant impact on their learning.

More detailed studies of students' homework habits are needed to fully understand the role that out of class practice plays in learning, but there is some evidence here that students are not benefiting from their practice. Despite the scaffolded reflection and in-class discussions about metacognition, most students did not see the value self-regulated learning. The one student who did, demonstrated a higher than expected performance on the FCI post-instruction test and the last in-class test. While the section's test average was comparable to prior ones', there likely was room for improvement if they had engaged in greater reflection.



\end{document}